\documentclass[pra,reprint,superscriptaddress]{revtex4-1}

\usepackage{amsmath, amsfonts,braket,graphicx,color}
\usepackage[colorlinks=true,citecolor=blue,linkcolor=magenta]{hyperref}

\global\long\def\avg#1{\langle#1\rangle}

\global\long\def\im{\imath}

\newcommand{\dg} {{\dagger}}
\newcommand{\pd} {{\phantom\dagger}}

\newcommand{\ci}[1] {{c_{#1}^{\pd}}}
\newcommand{\cid}[1] {c_{#1}^\dg}

\newcommand{\ai}[1] {a_{#1}^\pd}
\newcommand{\aid}[1] {a_{#1}^\dg}

\renewcommand{\Re} {\operatorname{Re}}
\renewcommand{\Im} {\operatorname{Im}}

\newcommand{\ql}{\mathcal{L}}
\newcommand{\qs}{\mathcal{S}}
\newcommand{\qr}{\mathcal{R}}

\newcommand{\qi}{\mathcal{I}}

\newcommand{\Dm}{D_\mathrm{max}}

\newcommand\trick[1]{}

\begin{document}

\title{Open System Tensor Networks and Kramers' Crossover for Quantum Transport}

\author{Gabriela W\'{o}jtowicz}
\affiliation{Jagiellonian University, Institute of Theoretical Physics, \L{}ojasiewicza 11, 30-348 Krak\'{o}w, Poland}
\author{Justin E. Elenewski} 
\affiliation{Biophysics Group, Microsystems and Nanotechnology Division, Physical Measurement Laboratory, National Institute of Standards and Technology, Gaithersburg, MD, USA}
\affiliation{Institute for Research in Electronics and Applied Physics, University of Maryland, College Park, MD, USA}
\author{Marek M. Rams}
\email{marek.rams@uj.edu.pl}
\affiliation{Jagiellonian University, Institute of Theoretical Physics, \L{}ojasiewicza 11, 30-348 Krak\'{o}w, Poland}
\author{Michael Zwolak}
\email{mpz@nist.gov}
\affiliation{Biophysics Group, Microsystems and Nanotechnology Division, Physical Measurement Laboratory, National Institute of Standards and Technology, Gaithersburg, MD, USA}

\begin{abstract}
Tensor networks are a powerful tool for  many--body ground states with limited entanglement.  These methods can nonetheless fail for certain time--dependent processes---such as quantum transport or quenches---where entanglement growth is linear in time.  Matrix-product-state decompositions of the resulting out-of-equilibrium states require a bond dimension that grows exponentially, imposing a hard limit on simulation timescales.  However, in the case of transport, if the reservoir modes of a closed system are arranged according to their scattering structure, the entanglement growth can be made logarithmic.  Here, we apply this ansatz to open systems via extended reservoirs that have explicit relaxation. This enables transport calculations that can access steady states, time dynamics and noise, and periodic driving (e.g., Floquet states). We demonstrate the  approach by calculating the transport characteristics of an open, interacting system. These results open a path to scalable and numerically systematic many--body transport calculations with tensor networks.
\end{abstract}

\maketitle

Many-body quantum systems are characterized by complex ground and dynamic states, reflecting the emergence of phenomena from superconductivity to exotic magnetism.  Furthermore, for nonequilibrium properties, systematic excitations can  yield a response that diverges markedly from the ground state~\cite{Eisert2015}.  The analytical treatment of many--body systems is thus challenging. This situation has driven the development of numerical methods such as quantum Monte Carlo~\cite{Foulkes2001,Wagner2016}, dynamical mean-field theory (DMFT)~\cite{Georges1996,Kotliar2006}, and tensor networks~\cite{Verstraete2008, Cirac2009,orus_review_2014, Hagegeman2016,Ran2018,Orus2019}, which now lie at the forefront of many--body theory.  Among these, tensor networks leverage the structure of correlations and entanglement to provide a local, numerically controllable many--body description---limiting computation to a submanifold of Hilbert space that captures the underlying state.

While tensor networks, such as matrix product states (MPSs), are extremely successful for correlated ground states, their application to time--dependent behavior can be stifled by a rapid growth of entropy~\cite{schuch_entropy_2008,eisert_entanglement_2013}.  In the context of quantum transport, this is due to scattering, which generates entangled electron--hole pairs in the adjacent contact regions~\cite{Rams2019}.  For a pair of electrodes, $\ql$ and $\qr$, held at a bias $\mu$, the attempt frequency for scattering events is $\mu / 2\pi$~\cite{Levitov1993}.  The bipartite entanglement entropy $S$ between them then grows linearly as $S \approx H[T(0)] \mu \, t / 2\pi$, where $H[T(\omega = 0)]$ is the binary entropy of the transmission probability $T(\omega)$ evaluated at the Fermi level (and $\mu$ is taken to be small)~\cite{Beenakker2006,Klich2009,chien_landauer_2014}.  Described qualitatively,  a simulation will fail when $S$ saturates the entanglement entropy permitted by a MPS with dimension $D$---a condition that is met when $S \approx \log_2 D$. Increasing the simulation duration requires an exponentially larger $D$ and thus exponentially larger computational requirements (see, e.g., Ref.~\onlinecite{Rams2019}).

\begin{figure}[b]
\vspace{-10pt}
\includegraphics[width=\columnwidth]{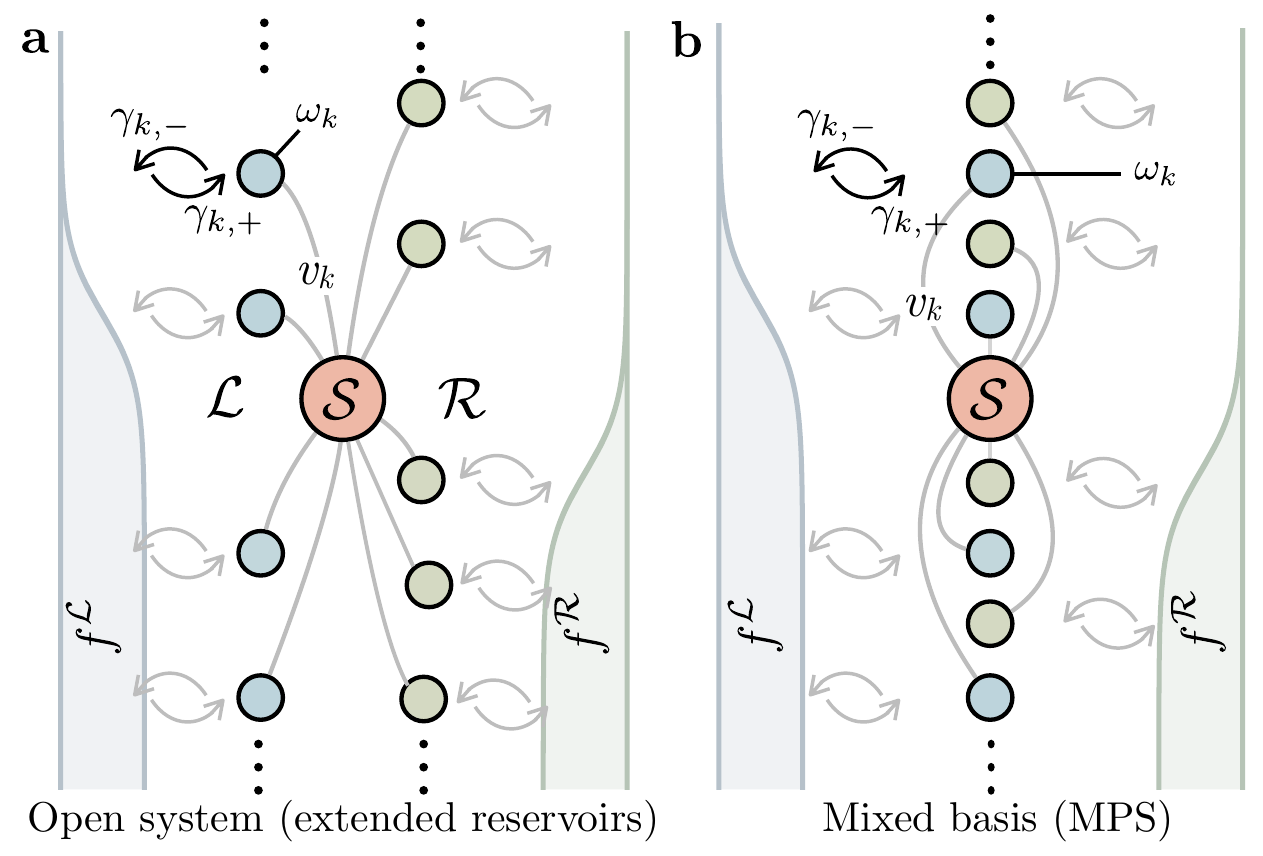}
\caption{{\bf Open system transport.}~{\bf a.}~An ``extended reservoir" approach for transport through a many--body system $\qs$. The system is flanked by explicit left, $\ql$, and right, $\qr$, reservoir modes of frequency $\omega_k$ and coupling $v_k$ to $\qs$. Implicit reservoirs relax the modes towards a Fermi-Dirac distribution via distinct injection ($\gamma_{k+}$) and depletion ($\gamma_{k-}$) rates tied to $f^{\ql/\qr}(\omega_k)$. {\bf b.}~The mixed energy--spatial basis has $\ql$ and $\qr$ modes arranged according to their energy $\hbar \omega_k$~\cite{Rams2019}.}~\label{fig:schematic}
\end{figure}

This exponential barrier can be mitigated by a basis that reflects the scattering structure of current--carrying states~\cite{Rams2019}: An incoming particle at energy $\hbar \omega_k$ will be transmitted from one reservoir, through the system, and into an outgoing state at the same energy (up to some characteristic spread in energy). A mixed energy--spatial basis, where (nearly) iso--energetic modes in $\ql$ and $\qr$ are adjacent on a one-dimensional (1D) lattice, captures the highly local entanglement structure in the energy domain. We recently leveraged this observation to develop a time--dependent MPS approach for closed systems. This approach shifts the temporal growth of entanglement from \emph{linear} to \emph{logarithmic}~\cite{Rams2019}.  The energy basis, also known in the literature as the ``star--geometry", has been employed in MPS impurity solvers for non-equilibrium DMFT~\cite{Wolf2014, bauernfeind_fork_2017} and in the study of quenched and Floquet states in the Anderson impurity model~\cite{He2017,He2019}, leading to suppression of entanglement~\cite{He2019}.  Nonetheless, our method is unique in pairing the reservoir modes according to the actual entanglement structure, giving accurate, extensive simulations of previously entanglement-limited non-equilibrium problems.

Here, we demonstrate that the mixed energy--spatial basis naturally permits open MPS simulations where implicit reservoirs offset carrier depletion in the explicit leads (reservoirs) $\ql$ and $\qr$ (Fig.~\ref{fig:schematic}). We first show that this method is numerically stable in the physical regime that reflects transport in the infinite reservoir limit (see Refs.~\onlinecite{Gruss2016, Elenewski2017}). We then apply our approach to a nontrivial example of transport through a two--site, interacting many--body impurity. The open system framework is critical for directly targeting steady states, including at finite temperature. In doing so, it provides a unified framework for addressing driven systems or those subject to time--dependent perturbations, such as Floquet dynamics and external noise on top of otherwise stationary states. These scenarios are generally difficult---and sometimes impossible---to access using existing methods.

Quantum transport is typically modeled using a composite system,  containing non--interacting left ($\ql$) and right ($\qr$) reservoirs that drive transport through an ``impurity" region (the system $\qs$; see Fig.~\ref{fig:schematic}a.)~\cite{Jauho1994}.  The Hamiltonian takes the form
\begin{equation}
H = H_\qs + H_\ql + H_\qr + H_\qi,
\end{equation}
where $H_\qs$ is the (many--body) Hamiltonian for $\qs$, $H_{\ql / \qr} = \sum_{k\in \ql\qr} \hbar \omega_k \aid{k} \ai{k}$ are the explicit reservoir Hamiltonians composed of $N_{\ql (\qr)}$ modes, and $H_\qi = \sum_{k\in \ql,\qr} \sum_{i\in \qs} \hbar v_{k,i} ( \cid{i} \ai{k} + \aid{k} \ci{i})$ is the interaction Hamiltonian that couples $\qs$ to $\ql\qr$.  The $\cid{i}$ ($\ci{i}$) and $\aid{k}$ ($\ai{k}$) are fermionic creation (annihilation) operators in $\qs$ and  $\ql\qr$, respectively.   We take the index $k$ to implicitly include relevant reservoir labels (state, spin, etc.), while $\omega_k$ and $v_{k,i}$ are the reservoir mode frequencies and system--reservoir coupling frequencies. 

While steady states can form if each reservoir contains an infinite number of explicit modes~\cite{Jauho1994}, only a finite reservoir can be simulated.  This will never give an actual steady state, making some parameter regimes and protocols (e.g., dynamic driving) difficult to access.  Implicit reservoirs offer a remedy, relaxing explicit modes to equilibrium distributions at different chemical potentials and/or temperatures~\footnote{There will still be a trade--off between relaxation time and accurate representation of long--time processes}. Simulating this requires an evolution for the density matrix $\rho$ of the $\ql\qs\qr$ composite system. A particularly useful approach is the Markovian master equation
\begin{align} \label{eq:fullMaster}
\dot{\rho} = - \frac{\im}{\hbar} [H, \rho]
    &+ \sum_k \gamma_{k+} \left( \aid{k} \rho \ai{k}
        - \frac{1}{2} \left \{ \ai{k} \aid{k}, \rho\right \}\right) \notag \\
    &+ \sum_k \gamma_{k-} \left( \ai{k} \rho \aid{k}
        - \frac{1}{2} \left \{ \aid{k} \ai{k}, \rho \right \} \right),
\end{align}
where $\{\cdot,\cdot\}$ is the anticommutator. The first term gives the  Hamiltonian evolution of $\rho$ (which includes explicit reservoirs), while the remaining terms inject and deplete particles into and from modes $k$ 
at a rate $\gamma_{k+}$ and $\gamma_{k-}$, respectively.
To ensure that explicit reservoirs relax to equilibrium in the absence of $\qs$, these rates are  $\gamma_{k+} \equiv \gamma f^\alpha (\omega_k)$ and $\gamma_{k-} \equiv \gamma [1 - f^\alpha (\omega_k)]$, where $f^\alpha (\omega_k)$ is the Fermi--Dirac distribution in the $\alpha \in \{\ql,\qr\}$ reservoir~\footnote{Here, $\gamma$ is the sole parameter controlling relaxation, but it can be made $k$--dependent at no cost}.

When the reservoirs are at different chemical potentials $\mu_\alpha$ (or temperatures), the bias $\mu = \mu_\ql - \mu_\qr$ will drive a current~\footnote{Since the bias is maintained externally by setting a chemical potential in the implicit reservoirs [via Lindblad terms in Eq.~\eqref{eq:fullMaster}], the applied bias does not appear in $\omega_k$. Thus, the mixed basis arranges the modes in $\ql$ and $\qr$ adjacent to each other according to their energies in the isolated reservoir Hamiltonians, thus following the natural mode structure of steady--state transport (see Fig.~\ref{fig:schematic}).}.  We call the reservoirs that are contained explicitly in the dynamics the \emph{extended reservoirs}.  This particular Markovian master equation has been widely employed (see Ref.~\onlinecite{Elenewski2017}) to describe transport in non--interacting systems~\cite{Dubi2009, dzhioev_super-fermion_2011, Zelovich2014, Zelovich2015, Zelovich2016, Zelovich2017, Ajisaka2015, Gruss2016, Hod2016, Elenewski2017, Gruss2017, Gruss2018}. It also follows naturally from the generic approach (explicit reservoir or bath states with  
Markovian broadening to represent the spectral function) suggested in Ref.~\cite{zwolak08-4} for open system dynamics. While it is provably correct in non--interacting and 
many-body cases~\cite{Gruss2016, Elenewski2017}, the extended reservoir size and $\gamma$ must lie in a certain regime~\cite{Gruss2016, Elenewski2017} to avoid unphysical relaxation artifacts. A related DMFT method uses an equation similar to Eq.~\eqref{eq:fullMaster}, but optimizes relaxation (including {\em intermode} relaxation) to represent the reservoir spectral function~\cite{Arrigoni2012,Dorda2014,dorda_auxiliary_2015,dorda_optimized_2017,Fugger2018}. While intermode relaxation can represent the spectral function more effectively, it increases the computational cost of the MPS simulation (we leave this for a later contribution).

In order to demonstrate the stability and behavior of tensor networks applied to the extended reservoir framework, we solve Eq.~\eqref{eq:fullMaster} with the time--dependent variational principle (TDVP)~\cite{Haegeman2011,Haegeman2016}, where the density matrix is vectorized to represent it within an MPS~\cite{zwolak_mixed-state_2004,MPDO_Cirac04}. While a variety of tensor network techniques exist \cite{Vidal2003b,*Vidal2004,Schmitteckert2004,zaletel_time-evolving_2015} we use TDVP as it is highly effective, 
accommodating a direct time evolution for any Hamiltonian or Lindbladian that may be represented as a matrix product operator (MPO). We note that time--evolving block decimation may also be efficacious~\cite{bauernfeind2019comparison} and leave open the question of an optimal implementation (via purification schemes~\cite{Werner_positive_2016,Jaschke_2018}, etc., see also Refs.~\onlinecite{Orus_2D_2017,Orus_open_2020} for a review). 
For details of our setup, refer to the note in ~\footnote{Our TDVP--based approach enforces operator Hermiticity by using real arithmetic and Hermitian operator basis to expand the MPS. To implement this method, we use the mapping of the master equation~\eqref{eq:fullMaster} onto a set of pseudospins using Jordan--Wigner transformation. The density matrix is then expanded using a local basis of Pauli matrices. The MPOs corresponding to super--operators $H \rho$ and $\rho H$ have a small bond dimension (six for the case of two-site impurity), and take on the same form as our prior closed system simulations (see, e.g., the Supplementary Material of Ref.~\onlinecite{Rams2019}), differing only by having superoperators instead of operators encoding the action of $H$ on the Pauli matrix basis. %
We note that both MPOs have a complex representation. However their sum (corresponding to $\im [H, \rho]$) can be made real using a suitable gauge transformation. We first add the two MPOs and compress the result, removing zero singular values. Then we put it into right canonical form and sweep the MPO from left to right. For a given MPO site $M_{l,r}^{m,n}$ (with virtual indices $\{l,r\}$) we consider the matrix $A = [\Re(M_{lmn, r}), \Im(M_{lmn, r})]$ concatenated along $r$ with dimension $D_r$. Calculating the singular value decomposition, $A=USV^\dag$, we take the last $D_r$ rows of $V$, forming the unitary matrix $X = V_{D_r+1:2 D_r, D_r+1:2 D_r} + \im V_{D_r+1:2 D_r, 1:D_r}$. The tensor $X$ provides a gauge transformation $X X^\dag$ which rotates $M_{lmn, r} X$ into the real form. The MPO of the Lindbladian term is also compact having bond dimension 3. During TDVP calculations, we restrict the MPS bond dimension $D_n$ at all cuts $n$ to be lower than a certain $D_\text{max}$. To account for the inhomogeneous nature of the setup and small correlations outside of the bias window we truncate  singular values $s\leq s_\text{min}$. In practice, we used $s_\text{min} = 10^{-6}$, which is sufficiently small to not affect the overall accuracy significantly (it is predominantly controlled by $D_\text{max}$). This involves a hybrid one-site/two-site TDVP scheme, where the two-site scheme is employed only if it is necessary to enlarge the bond dimension $D_n$. To solve the local evolution problems for non-Hermitian effective ``Hamiltonians'' appearing in the TDVP algorithm, we employ the Krylov-based procedure of Ref.~\onlinecite{Niesen_Krylov}. Finally, in order to avoid the issue of the strong correlations between impurity sites we merge those into a single one.}.

Our approach uses the reservoirs' energy (or momentum) bases and thus there is, in principle, {\em no requirement} for their spatial dimensionality: They can be of arbitrary dimension and have long--range hopping both within the reservoirs and to $\qs$. 
The reservoirs only need to be non--interacting, without a direct coupling between left and right regions (both requirements can be relaxed~\cite{Rams2019}, albeit with a loss of efficiency). As an example, we consider a system composed of two sites with the Hamiltonian
\begin{equation} \label{eq:HS}
H_{\qs} = \hbar v_S (\cid{1} c_{2} + \cid{2} c_{1}) + \hbar U n_1 n_2 ,
\end{equation}
where $n_i$ the number operator at site $i$, $U$ is the interaction strength, and $N_\qr = N_\ql = N$ spinless reservoir modes defined by frequencies
$\omega_k=2\omega_0 \cos (k \pi /(N+1))$ and couplings $v_{k,i}=v \sqrt{2/\left(N+1\right)} \sin\left(k\pi/\left(N+1\right)\right)$ for $(k,i)\in \{(\ql,1), (\qr,2)\}$, where $k$ carries a numerical (1 to $N$) and reservoir label. This is the single-particle eigenbasis for spatially one-dimensional reservoirs, with the left reservoir connected only to site 1 of $\qs$ and the right connected to site 2 of $\qs$, of hopping frequency $\omega_0$. Physically, this is the time--independent version of a model for photoconductive molecular devices where spin plays no role~\cite{Dubi_2018}.

\begin{figure}[h!]
\includegraphics[width=\columnwidth]{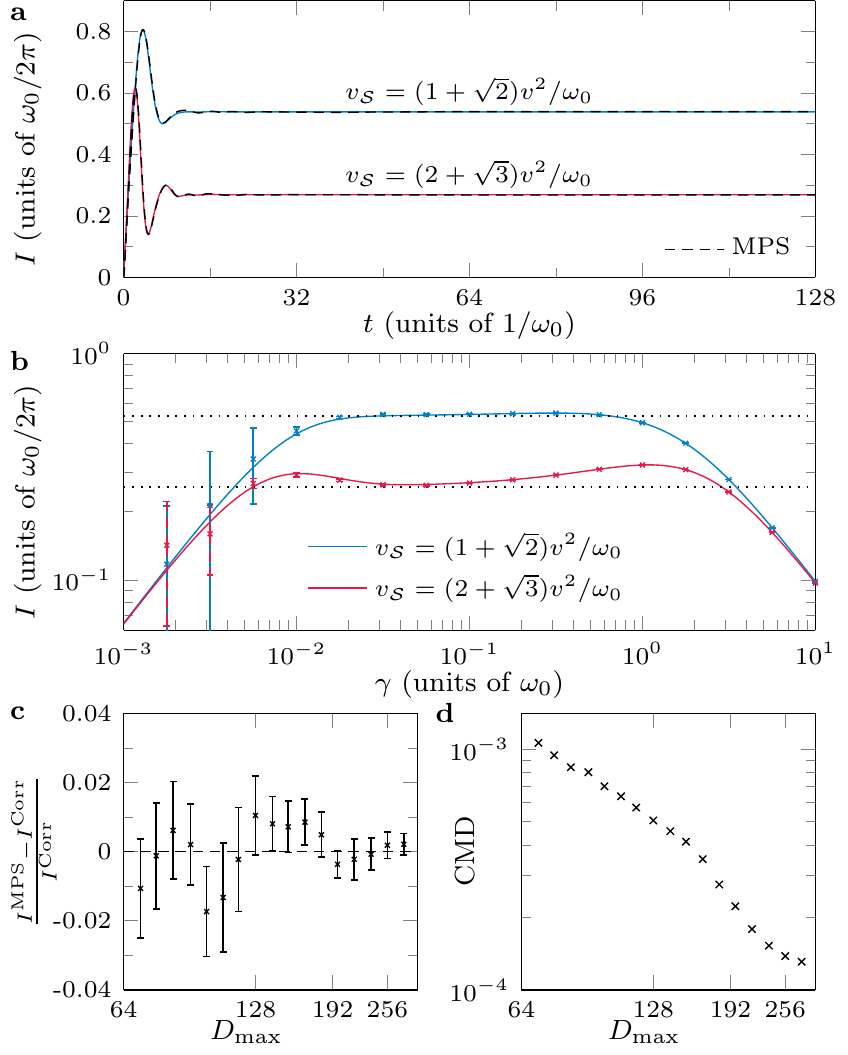}
\vspace{-10 pt}
\caption{{\bf Non--interacting benchmark.}~{\bf a.}~Transient current for a non--interacting ($U=0$), spinless, two--site system.   Data are given for two values of $v_S$ at fixed relaxation $\gamma = 0.1\,\omega_0$.
Mixed basis MPS dynamics (black, dashed line) match the exact result (solid, colored lines) even at a modest $\Dm=256$. {\bf b.}~Steady--state current $I(t=\infty)$ versus  $\gamma$ from diagonalizing the exact Lindbladian (solid lines) and the long--time limit, up to time $t_\text{max}$, of MPS propagation (data points).  Exact and MPS data compare well, with deviations only in the small $\gamma$ regime. The currents for infinite reservoirs without relaxation, computed using non-equilibrium Green's functions~\cite{Jauho1994,Gruss2016}, are shown as dotted lines. {\bf c.}~Relative deviation of the current $(I^{\text{MPS}} - I^{\text{Corr}})/I^{\text{Corr}}$ and {\bf d.}~relative trace distance (CMD) at the last time of the simulation between the MPS and exact results as a function of $\Dm$ (the relative trace distance divides the trace distance by the trace of the exact correlation matrix). Panels {\bf c} and {\bf d} show the convergence in $\Dm$ at $\gamma = 0.1\,\omega_0$ and $v_S=(1+\sqrt{2})v^2/\omega_0$. All calculations have a bias of $\mu = \omega_0/2$ across implicit reservoirs of $N = 128$ modes each, starting from Fermi--Dirac occupations ($k_B T = \hbar \omega_0 / 40$), and a system--reservoir coupling $v = \omega_0/2$. The steady-state current from MPS simulations is taken as an average over the last  $\Delta t = t_\text{max} / 10$. The error bars are $\pm \sigma$ with $\sigma^2=\sigma_1^2+\sigma_2^2$, where $\sigma_1$ is the fluctuations of $I$ in $\Delta t$ and $\sigma_2^2 = \sum_i |I_i-I|^2/3$ is the mismatch of currents at different interfaces $i$. The latter likely captures the bulk of the error from truncating the bond dimension $D$---the dominant source of errors here.
}\label{fig:mps1}
\end{figure}

Figure~\ref{fig:mps1} shows benchmark calculations of the current at $U=0$ that compare the exact solution to the MPS simulations~\footnote{The initial state in all simulations is the steady state of Eq.~\eqref{eq:fullMaster}, with $\qs$ half full, without $H_\qi$ in $H$.}. We calculate the current as the average across three interfaces $I=\avg{I_i}$ with $i \in \{{\ql \qs}, {\qs \qs}, {\qr \qr} \}$. Where at the $\ql\qs$ interface, $I_{\ql\qs}(t) = -4 \sum_{k \in \ql} v_k \, \text{Im} \langle \aid{k} c_{1}\rangle$ with similar notation for the others~\footnote{An extra factor of $2$ is included to account for spin}. In the steady state, these currents should be identical and constant. Thus, their variation in time and variance in $i$ quantify errors within the calculation. 

The transient current $I(t)$, in Fig.~\ref{fig:mps1}a, rapidly reaches a steady state at intermediate reservoir relaxation, and is in tight correspondence with simulations that employ the exact correlation matrix. The required $D$ is also modest and time evolution does not break down like closed-system approaches in the spatial basis (see Ref.~\cite{Rams2019}). Rather, once in the steady state, the current remains constant up to small, easily quantifiable errors. Prior to reaching the steady state, there are oscillations that are a manifestation of the Gibbs phenomenon~\cite{zwolak_communication:_2018} and its interplay with the initial state and other interactions. The rise time to a quasisteady state is set by the reservoir bandwidth ($\propto 1/\omega_0$), but ultimately the time to reach the steady state---where all oscillations, which can be both persistent~\cite{Rams2019,branschadel_conductance_2010} and algebraically decaying~\cite{zwolak_communication:_2018}, disappear---will be set by the  relaxation time $\gamma^{-1}$.

The relaxation, however, also gives rise to distinct regimes of conduction, as seen in Fig.~\ref{fig:mps1}b~\footnote{To help exhaustively map the transport regimes, we introduce a protocol where the terminal MPS from a given TDVP evolution is the starting state for a successively smaller relaxation $\gamma_m$.  At each stage, we adopt a timestep $\Delta t_m = \min [1, 1/\gamma_m]$ and maximal simulation time $t_{\max} = 10  \Delta t_m \max[1 /\gamma_m, 100]$ which are defined in terms of $\gamma_m$, accelerating convergence to the steady state. The steady state can also be targeted directly with the density matrix renormalization group~\cite{cui2015variational}}: The current initially increases linearly with $\gamma$, plateaus, and then decreases as $1/\gamma$~\footnote{There are exceptions to the $\gamma$ and $1/\gamma$ regimes, depending on energy gaps and symmetries, see Ref.~\onlinecite{Gruss2017}.}. The three distinct regimes reflect a simulation analog of Kramers' turnover in the rate of condensed-phase reactions~\cite{kramers1940brownian}. Instead of a turnover versus the solution friction, the electronic conductance depends on the external relaxation~\cite{Gruss2016,Elenewski2017,Gruss2017}, which is a dependence also found in thermal transport~\cite{Velizhanin2011,Chien2013,Velizhanin2015,Chien2017,Chien2018}. In this context, the steady state at small $\gamma$ resembles a reactive system that is rate--limited by how quickly equilibrium is restored among reactants after a subset of them proceeds to products. For transport, this gives a linear dependence in $\gamma$ (and $N$), as this gives the rate at which particles and holes are replenished in the reservoirs.
At large $\gamma$, transport  parallels reactions that are controlled by rapid environmental processes (e.g., solvent dynamics), which redirect intermediates along the reactive pathway back to reactants. For transport, the rapid relaxation suppresses the development of coherence, which is a necessary condition for current to flow. The intrinsic rate,  when friction and relaxation are absent, is the dominant factor between small and large $\gamma$ limits. Physical behavior corresponding to infinite, relaxation--free reservoirs  occurs only in this intermediate regime~\cite{Gruss2016,Elenewski2017}.

In addition to performing simulations for intermediate relaxation, there are additional requirements to properly represent transport. Markovian relaxation is inherently {\em not physical}~\cite{Gruss2016,Elenewski2017}, as it occupies modes according to their frequencies in isolation rather than those properly broadened by contact with the implicit reservoirs. Thus, $\gamma$ must be sufficiently lower than the thermal broadening, $\gamma \ll k_B T / \hbar$. The current should also be on the plateau, which occurs at $\gamma \approx W / N$, where $W$ is the reservoir bandwidth ($W = 4\omega_0$ in our example). This requires that $N \gg \hbar W / k_BT$ for physical simulations~\cite{Gruss2016,Elenewski2017}.  
In practice, we take values of $\gamma$ to be on the plateau for a given $N$, and compute errors due to the variation of $N$ and $\gamma$. The MPS simulations work very well from large $\gamma$ down to the plateau edge, Fig.~\ref{fig:mps1}b. As $\gamma$ leaves the plateau ($\gamma<W/N$), long--time coherence of particles flowing back and forth in the $\ql\qs\qr$ system make MPS simulations difficult, albeit in a turnover regime that is uninteresting from both physical and practical standpoints~\footnote{Off the plateau, $\gamma<W/N$, correlations are more moderate than in the physical regime but also more diffuse (relative to their maximum in the bias window). This regime is also complicated by the timescales needed to reach the steady state and the weak current being buried in the background (i.e., difficult to extract). Nonetheless, truncation dominates the error for small $\gamma$.}. 
The error in the current, arising from MPS truncation, decays steadily, but non--monotonically, with $D$ in the plateau regime, see Fig.~\ref{fig:mps1}c. The convergence of the state itself is monotonic in $D$ (truncation gives the dominant source of error at fixed $N$ and $\gamma$). We quantify this using the relative trace distance between the exact and MPS-based single-particle correlation matrices, see Fig.~\ref{fig:mps1}d.

\begin{figure}
\includegraphics[width=\columnwidth]{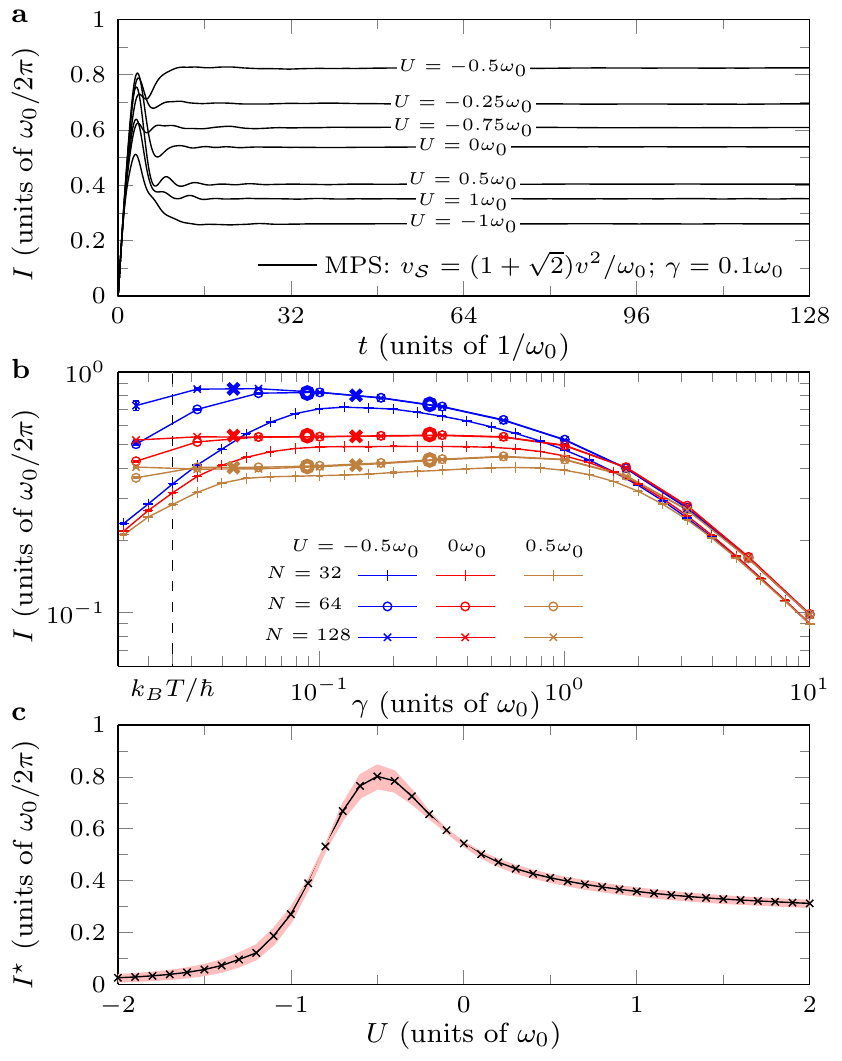}
\vspace{-10pt}
\caption{{\bf Transport in an interacting, two--site system.}~{\bf a.}~Current versus time showing the initial transient oscillations and the rapid approach to the steady state for several $U$ and $N=128$ with $\Dm=256$. {\bf b.}~Steady--state current versus $\gamma$ for several values of $U$ and at $N=32$, $64$ and $128$ and $\Dm=256$. This demonstrates the same regimes as the non--interacting case. Simulations are truncated at $\gamma = 0.015 \,\omega_0$, as converging this unphysical regime to the same accuracy as for larger $\gamma$ is exceptionally costly. {\bf c.}~Steady--state current in the thermodynamic limit, $I^\star$, versus the many--body coupling $U$ using $N=128$, $\gamma^\star=10^{-3/4}\, 32 \omega_0/N,$
and $D_{max}=256$. All parameters are identical to those in Fig.~\ref{fig:mps1} unless otherwise indicated. The error bars in {\bf b} are the same as in Fig.~\ref{fig:mps1}. The error bands in {\bf c} also include error due to improper plateau formation, $\sigma_3^2$, which is the variance of the four bold data points in {\bf b} (at $\gamma=10^{-1/4}\,32 \omega_0 /N$ and $10^{-3/4}\,32 \omega_0/N $ for both $N=64$ and 128).} \label{fig:mps2} 
\end{figure}

The well-behaved nature of the simulations carries over to the many-body case. Figure~\ref{fig:mps2}a shows $I(t)$ for several values of the interaction strength $U$. Just as with $U=0$, the current rapidly approaches its steady state and remains there. A modest $D$ converges the results regardless of $U$. Figure~\ref{fig:mps2}b shows the current versus $\gamma$, demonstrating the existence of a many--body simulation analog to Kramers' turnover. This behavior is expected, as the junction plays only a tangential role in the mechanism of transport in the small- and large-$\gamma$ regimes. Nonetheless, the many-body interaction does influence convergence to a smooth plateau. Moreover, the small-$\gamma$ regime remains difficult for MPS but does take on a linear relationship where $I \propto \gamma$.

Using the Kramers' turnover in Fig.~\ref{fig:mps2}b, we can find the best estimator, $I^\star$, for the current in the thermodynamic limit, and estimate its error. We take $I^\star$ to be at  $\gamma^\star=10^{-3/4}\,32\omega_0/N$, which is on the plateau just before the turnover to the small $\gamma$ regime. Error estimates incorporate the effect of a finite bond dimension (see $\sigma_1$ and $\sigma_2$ in Fig.~\ref{fig:mps1}) and plateau formation (see $\sigma_3$ in Fig.~\ref{fig:mps2}). In this case, incomplete plateau formation is the main contribution to the error. Figure~\ref{fig:mps2}c shows $I^\star$ versus $U$, which  exhibits an enhanced conductance at $U=-0.5\,\omega_0$. There are two eigenstates in $\qs$ at $\pm v_\qs$ when $U=0$. When $U$ is sufficiently large and negative, there is a bound state of two fermions in $\qs$ lying outside the bias window, blocking the current (i.e., it will decay to zero as $U\to -\infty$). When $U$ approaches but remains less than zero, the system mode at $- v_\qs$ is nearly occupied, and the higher-energy mode is effectively pulled down in energy by the many-body interaction. Neglecting the lower energy mode, other than a mean-field effect on the higher energy mode, gives a peak current at $U=-4v_\qs/3$. This is in reasonable agreement with the full many-body result. As $U$ becomes repulsive, the system will begin to have only a single particle present. Nevertheless, this can sustain a finite current (half the current when $U=0$) even as $U\to\infty$, giving rise to the asymmetry between very attractive and very repulsive interactions.

\begin{figure}[t]
\includegraphics[width=\columnwidth]{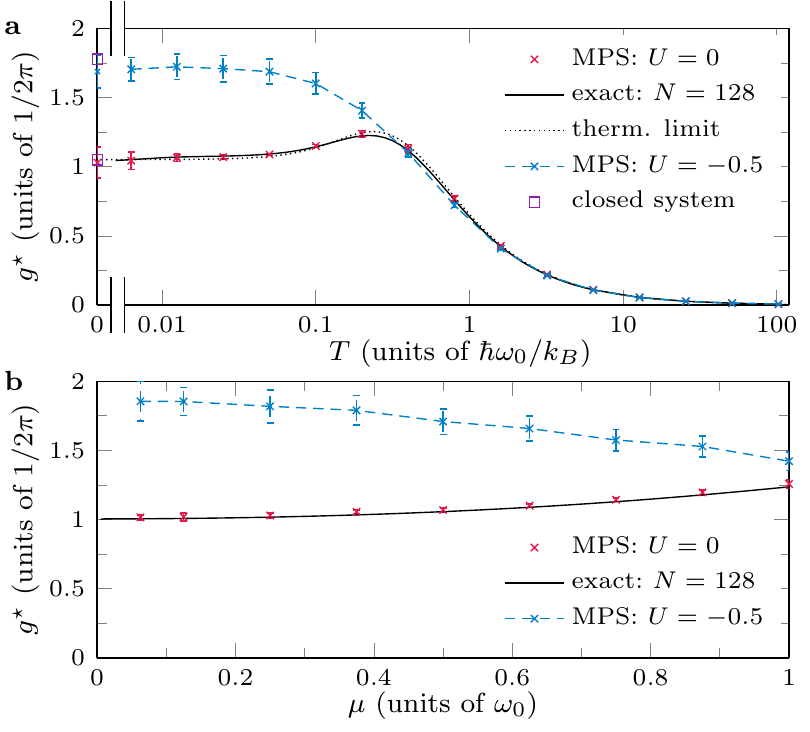}
\vspace{-10pt}
\caption{{\bf Conductance versus temperature and bias.}~{\bf a.}~Conductance versus the temperature $T$ at $\mu = \omega_0 / 2$. Closed--system calculations follow Ref.~\onlinecite{Rams2019} and use $N=128$. {\bf b.}~Conductance versus bias $\mu$ at $k_B T= \hbar \omega_0/40$. The setup for the simulations, and the determination of $I^\star=\sigma^\star \mu$ and its error, are identical to Fig.~\ref{fig:mps2}c.}
\label{fig:mps3}
\end{figure}

The approach can be employed in both linear response and out-of-linear response, at finite temperature, and for time-dependent processes (e.g., Floquet states or noise). Figure~\ref{fig:mps3}a shows the conductance $\sigma^\star$ across a range of temperatures and biases for both interacting and noninteracting cases. There is a robust correspondence in $\sigma^\star$ among extended reservoir MPS, exact correlation matrix simulations, and the thermodynamic limit from non--equilibrium Green's functions~\cite{Jauho1994,Gruss2016}. 
The magnitude of error becomes significant only when $T$ is near zero as in Fig.~\ref{fig:mps3}a. Even then, it is suitable for practical calculations. There is also good agreement with closed-system MPS at $T=0$~\cite{Rams2019}, with the latter being within the error estimate, confirming that our protocol is reliable. To fully approach the thermodynamic limit, $N$ should be enlarged and $\gamma$ should be diminished. Nonetheless, these data indicate that our open--system method may be applied across a broad range of parameters with no modification.

While MPS approaches have previously been employed to study transport in real--time for spatially 1D models~\cite{cazalilla_time-dependent_2002, zwolak_mixed-state_2004,al-hassanieh_adaptive_2006,Schmitteckert06-1,schneider_conductance_2006,dias_da_silva_transport_2008,heidrich-meisner_real-time_2009,branschadel_conductance_2010,chien_interaction-induced_2013,gruss_energy-resolved_2018}, with a numerical renormalization group approach to the reservoirs~\cite{schwarz_nonequilibrium_2018}, and in linear response~\cite{bohr_dmrg_2006,bohr_strong_2007}, our mixed--basis approach eliminates constraints on accessible time and spatial scales while also not requiring a one-dimensional, or quasi-one-dimensional, lattice for the noninteracting reservoirs~\cite{Rams2019}. The open system method here enables the direct computation of steady-state currents, making their determination comparable to finding ground states (i.e., finding a stationary state). Our approach will also be useful for finding Floquet states and studying the effect of artificial gauge fields, examining time--dependent processes that perturb the system around its stationary state, handling transport when long timescales appear (e.g., due to many-body interactions), and giving the right framework for coarse--graining reservoir modes~\cite{Zwolak08-2}, all at a finite temperature that will help in the study of thermoelectrics and determining many--body temperature scales. It may also help in solving other boundary driven problems (e.g., 1D lattices driven by Markovian processes only at the ends). Overall, this scheme will permit the accurate simulation of challenging many--body systems, where larger and more complex impurity physics gives  rise to intricate behavior. 

\begin{acknowledgments}
{\it Note added}: Recently, related work appeared that implemented the extended reservoir approach with MPS simulations but with different underlying machinery~\cite{brenes_tensor-network_2019, Lotem_tensor-network_2020}.

J.E.E. acknowledges support under the Cooperative Research Agreement between the University of Maryland and the National Institute for Standards and Technology Physical Measurement Laboratory, Award No. 70NANB14H209, through the University of Maryland. We acknowledge support from the National Science Center, Poland under Projects No. 2016/23/B/ST3/00830 (G.W.) and No. 2016/23/D/ST3/00384 (M.M.R.).
\end{acknowledgments}

\end{document}